# Community detection for networks with unipartite and bipartite structure

**Chang Chang[1,2] and Chao Tang[2]**

[1] School of Life Sciences, Peking University, Beijing 100871, People's Republic of China
[2] Center for Quantitative Biology and Peking-Tsinghua Center for Life Sciences, Peking University, Beijing 100871, People's Republic of China
E-mail: chang.connected@pku.edu.cn and tangc@pku.edu.cn

## Abstract

Finding community structures in networks is important in network science, technology, and applications. To date, most algorithms that aim to find community structures only focus either on unipartite or bipartite networks. A unipartite network consists of one set of nodes and a bipartite network consists of two nonoverlapping sets of nodes with only links joining the nodes in different sets. However, a third type of network exists, defined here as the mixture network. Just like a bipartite network, a mixture network also consists of two sets of nodes, but some nodes may simultaneously belong to two sets, which breaks the nonoverlapping restriction of a bipartite network. The mixture network can be considered as a general case, with unipartite and bipartite networks viewed as its limiting cases. A mixture network can represent not only all the unipartite and bipartite networks, but also a wide range of real-world networks that cannot be properly represented as either unipartite or bipartite networks in fields such as biology and social science. Based on this observation, we first propose a probabilistic model that can find modules in unipartite, bipartite, and mixture networks in a unified framework based on the link community model for a unipartite undirected network [B Ball *et al* (2011 *Phys. Rev.* E **84** 036103)]. We test our algorithm on synthetic networks (both overlapping and nonoverlapping communities) and apply it to two real-world networks: a southern women bipartite network and a human transcriptional regulatory mixture network. The results suggest that our model performs well for all three types of networks, is competitive with other algorithms for unipartite or bipartite networks, and is applicable to real-world networks.

Online supplementary data available from stacks.iop.org/NJP/16/093001/mmedia

## 1. Introduction

Community structure, or network modules, has become a fruitful topic in fields such as physics, mathematics, biology, and social science [1]. To date, most algorithms focusing on this problem operate on either unipartite [2–7] or bipartite networks [8–10] and some can handle both [11, 12] (reviewed in [1, 13]). A unipartite network consists of a vertex set and an edge set that join pairs of vertices. A bipartite network consists of two disjointed sets of vertices and a set of edges in which each edge only joins vertices in different sets. Most complex networks in nature and society are represented as either unipartite or bipartite [1].

However, a third type of network exists, defined here as the mixture network. Just like a bipartite network, a mixture network consists of two sets of nodes; however, some nodes may simultaneously belong to two sets rather than one, which breaks the nonoverlapping restriction. When all nodes belong to two sets, a mixture network is unipartite; when each node only belongs to one of the two sets, it is bipartite; when only a part of the nodes belongs to two sets, it is neither unipartite nor bipartite. Thus, the mixture network can be considered as a general case, with unipartite and bipartite networks viewed as limiting cases. A mixture network can represent not only all the unipartite and bipartite networks, but also a wide range of real-world networks that cannot be properly represented as unipartite or bipartite networks in fields such as biology and social science (for convenience of illustration, we use only a mixture network to denote the mixture networks that are neither unipartite nor bipartite hereafter):

  (i) Transcriptional regulatory networks [14–16]: one vertex set represents transcriptional factors (TFs), which play the role of regulators, and another set represents a collection of downstream target genes. Because some TFs can also play the role of target genes, interactions can occur not only between a TF and a target gene, but also between TFs [14], suggesting that certain TFs belong to both vertex sets simultaneously rather than just one. Considering that most target genes do not play the role of regulators, edges exist between the two vertex sets that are neither disjointed nor identical. Consequently, transcriptional regulatory networks are mixture networks rather than unipartite or bipartite networks.

 (ii) Phosphorylation networks [17]: one vertex set consists of kinases or phosphatases, and another set consists of target proteins that are being phosphorylated or dephosphorylated. Some kinases or phosphatases can also be phosphorylated or dephosphorylated by other kinases or phosphatases, playing both roles in the network.

(iii) Shareholding networks [18]: one vertex set represents the owners and another set represents corporations or stocks. Because a corporation can also be owned by another corporation, and not all owners are corporations, this means shareholding networks are examples of mixture networks.

(iv) Some genetic interaction networks built from epistatic mini-array profiles [19–22]: in such networks, one vertex set represents query genes and the other set represents array genes. Two genes are linked if they have a genetic interaction. One gene can be either a query gene or an array gene, and sometimes it can be both. Ideally, the case of measuring genetic interactions between all gene pairs would lead to a unipartite network [23, 24]. Practically, not all pairs are measured because of research goals or funding limitations in many studies, such as in [19–21], or in the epistatic mini-array profiles that focus on the cell cycle of *Saccharomyces cerevisiae* and *Schizosaccharomyces pombe* conducted by our group [22], so that only interactions between two subsets of genes that share common genes are measured.

(v) Some protein interaction networks built by the yeast two-hybrid method [25]: one vertex set represents bait proteins and another set represents prey proteins. The reason the network in [25] is a mixture network is because of its experimental design.

The preceding examples can be summarized into two types of mixture networks. The first type consists of two vertex sets of two roles, in which some nodes can play both roles, thus they simultaneously belong to two sets rather than one. The second type is a subset of a corresponding unipartite network, because only links between two subsets of nodes that share certain common elements are known. All examples mentioned previously belong to the first type and examples (iv–v) belong to the second type.

In previous studies, different approaches have been used to analyze mixture networks, although the concept is not explicitly formed. However, many approaches suffer from drawbacks when used in mixture networks without considering their own characteristics. First, directly using algorithms that can work for bipartite networks, such as biclustering [26] or hierarchical clustering [20], will omit the association between pairs of nodes that are actually one. Such nodes may be partitioned into different modules improperly, even when a community detection algorithm does not allow for an overlapping community or when they are assigned to the same module with two different weights or probabilities. In addition, projecting a mixture network into a unipartite network is an alternative choice [21], but information will be lost in such a process, as it would be for a bipartite network [8, 27], and only one vertex set, rather than both, can be assigned to modules following the transformation. Furthermore, imputation methods can be used to convert the second type of mixture networks to unipartite networks by predicting missing values [28]. However, the performance of a community detection algorithm will then depend on the performance of the imputation method used. Thus, models specific to mixture networks are necessary.

We aimed to find network modules in unipartite, bipartite, and mixture networks in a unified frame that would allow the relationship between two vertex sets to be identical, disjointed, or neither. To achieve this goal, we propose a probabilistic method that operates on all three types of networks based on the link community model for a unipartite undirected network developed by Ball *et al* [6] (denoted as BKN hereafter).

This paper is organized as follows. First, we review BKN for a unipartite undirected network in section 2.1. We redefine this model in the context of a bipartite network in section 2.2 and generalize it to a mixture network (the general case) in section 2.3. In sections 3 and 4, we explore the performance of the model on synthetic networks generated using the model itself and those generated by sampling, which are two ways of generating all three types of networks. In sections 5 and 6, we apply the model to random unipartite directed networks and bipartite networks. Subsequently, we apply the model to real-world data of bipartite and mixture networks in section 7. Note that simulations and applications for real-world data solely for unipartite undirected networks do not need to be discussed in this paper because they have been addressed in [6]. In section 8, we provide a discussion.

## 2. Generative models for module detection

### 2.1. Unipartite network

Because our model is an extension and generalization of BKN, we briefly review this generative model for a unipartite undirected network in this subsection.

Let $G_U = (V, E)$ denote a unipartite undirected network, where $V$ is the vertex set and $E$ is the edge set. Here, **A** represents the adjacency matrix, where $A_{ij} = 1$ if there is a link between vertex $i$ and vertex $j$, and $A_{ii} = 2$ if there is a self-loop of vertex $i$. Further, suppose the number of modules $K$ is given. BKN is parameterized by a set of parameters $\theta$, such that $\theta_{iz}$ represents the propensity of vertex $i$ to belong to module $z$. Specifically, the physical meaning of $\theta_{iz}\theta_{jz}$ is the expected number of links between vertices $i$ and $j$ belonging to module $z$, with the exact number being Poisson distributed. By summing over all $K$ modules, $\sum_z \theta_{iz}\theta_{jz}$ represents the expected number between vertices $i$ and $j$, the exact number of which is also Poisson distributed, because the sum of independent Poisson variables is still Poisson distributed. Given that the actual number of edges between $i$ and $j$ is known, the likelihood function of the unipartite undirected network $G_U$ with the adjacency matrix **A** is

$$P\left(G_U \middle| \theta\right) = \prod_{i<j} \frac{\left(\sum_z \theta_{iz}\theta_{jz}\right)^{A_{ij}}}{A_{ij}!} \exp\left(-\sum_z \theta_{iz}\theta_{jz}\right)$$

$$\times \prod_i \frac{\left(\frac{1}{2}\sum_z \theta_{iz}\theta_{iz}\right)^{A_{ii}/2}}{\left(A_{ii}/2\right)!} \exp\left(-\frac{1}{2}\sum_z \theta_{iz}\theta_{iz}\right). \tag{1}$$

BKN can be fitted to an observed network according to the maximum likelihood principle, with respect to parameters $\theta_{iz}$. By taking the logarithm of equation (1), and discarding constant terms, the log likelihood function of the model can be written as

$$\ln P\left(G_U \middle| \theta\right) = \sum_{ij} A_{ij} \ln\left(\sum_z \theta_{iz}\theta_{jz}\right) - \sum_{ijz} \theta_{iz}\theta_{jz}. \tag{2}$$

Applying Jensen's inequality, we obtain

$$\ln P\left(G_U \middle| \theta\right) \geqslant \sum_{ijz} \left[A_{ij} q_{ij}(z) \ln\left(\frac{\theta_{iz}\theta_{jz}}{q_{ij}(z)}\right) - \theta_{iz}\theta_{jz}\right] \tag{3}$$

by introducing an arbitrary variable $q_{ij}(z)$, which satisfies $\sum_z q_{ij}(z) = 1$. When

$$q_{ij}(z) = \frac{\theta_{iz}\theta_{jz}}{\sum_z \theta_{iz}\theta_{jz}}, \qquad (4)$$

the inequality becomes an exact equality. By differentiating equation (3), with respect to $\theta_{iz}$, we obtain

$$\theta_{iz} = \frac{\sum_j A_{ij} q_{ij}(z)}{\sum_j \theta_{jz}}, \qquad (5)$$

which gives the optimal value of $\theta_{iz}$. In [6], Ball *et al* further change the form of equation (5) to

$$\theta_{iz} = \frac{\sum_j A_{ij} q_{ij}(z)}{\sqrt{\sum_{ij} A_{ij} q_{ij}(z)}}, \qquad (6)$$

using the fact that $\sum_i \theta_{iz} = \sum_j \theta_{jz}$.

BKN is an expectation–maximization (EM) algorithm, which can be used to find the maximum likelihood in an iterative manner. The log likelihood increases monotonically by updating the parameters iteratively using equations (4) and (5) or equations (4) and (6), which can be viewed as two variants of EM—the former is known as the expectation conditional maximization (ECM) [29], and the latter as an incremental variant of the EM [30]. Multiple random initializations are required to escape from local maxima and the division with the highest log likelihood is chosen as the result.

Because the physical meaning of $q_{ij}(z)$ is the weight of the edge between vertices $i$ and $j$ belonging to module $z$, this quantity is used to infer link communities in the network. Mathematically, vertex $i$ belongs to module $z$ if $\sum_j A_{ij} q_{ij}(z) \geqslant 1$ [3], and the physical meaning of the left-hand side of this inequality is the average number of links belonging to module $z$ that link to vertex $i$.

## 2.2. Bipartite network

In this subsection, we redefine BKN in the context of a bipartite network. In a unipartite network, all links exist within the same group of vertices. Consequently, one set of parameters, $\theta$, is enough to describe the underlying structure of a unipartite network. On the other hand, there are two sets of vertices in a bipartite network; thus, two sets of parameters should be used for the parameterization. In a bipartite network $G_B = (U, V, E)$, $U$ and $V$ denote two disjointed vertex sets and $E$ denotes the edge set. We use two sets of parameters $\theta^{(U)}$ and $\theta^{(V)}$ to describe the underlying structure of a bipartite network, such that $\theta_{iz}^{(U)}$ represents the propensity of vertex $i$ in the vertices of set $U$ belonging to module $z$, and $\theta_{jz}^{(V)}$ represents the propensity of vertex $j$ in the vertices of set $V$ belonging to module $z$. Similar to BKN, the physical meaning of $\theta_{iz}^{(U)}\theta_{jz}^{(V)}$ is the expected number of links between vertex $i$ in vertex set $U$ and vertex $j$ in vertex set $V$ belonging to module $z$, with the exact number being Poisson distributed. By summing over all $K$ modules, $\sum_z \theta_{iz}^{(U)} \theta_{jz}^{(V)}$ represents the expected number of links between vertex $i$ in vertex set $U$ and vertex $j$ in vertex set $V$, the exact number of which remains Poisson distributed. The likelihood function of generating a bipartite network $G_B$ is

$$P\left(G_B \middle| \theta^{(U)}, \theta^{(V)}\right) = \prod_{ij} \frac{\left(\sum_z \theta_{iz}^{(U)} \theta_{jz}^{(V)}\right)^{A_{ij}}}{A_{ij}!} \exp\left(-\sum_z \theta_{iz}^{(U)} \theta_{jz}^{(V)}\right). \tag{7}$$

---

[3] In Ball *et al* this criterion is defined using strict inequality. However, in our implementation, the equal condition (or nearly equal) is also considered. Suppose there is a node with only one link, which belongs to module $z$ with weight very close to 1; it is reasonable to assign this node to module $z$.

By taking the logarithm of the likelihood function, and applying Jensen's inequality with an arbitrary variable $q_{ij}(z)$ that satisfies $\sum_z q_{ij}(z) = 1$, we obtain

$$\ln P\left(G_B \middle| \theta^{(U)}, \theta^{(V)}\right) \geqslant \sum_{ijz} \left[A_{ij} q_{ij}(z) \ln\left(\frac{\theta_{iz}^{(U)} \theta_{jz}^{(V)}}{q_{ij}(z)}\right) - \theta_{iz}^{(U)} \theta_{jz}^{(V)}\right], \tag{8}$$

where the optimal value of $q_{ij}(z)$ is given by

$$q_{ij}(z) = \frac{\theta_{iz}^{(U)} \theta_{jz}^{(V)}}{\sum_z \theta_{iz}^{(U)} \theta_{jz}^{(V)}}, \tag{9}$$

with the physical meaning unchanged compared to that of BKN. The optimal values of $\theta_{iz}^{(U)}$ and $\theta_{jz}^{(V)}$ are given by

$$\theta_{iz}^{(U)} = \frac{\sum_j A_{ij} q_{ij}(z)}{\sum_j \theta_{jz}^{(V)}}, \qquad \theta_{jz}^{(V)} = \frac{\sum_i A_{ij} q_{ij}(z)}{\sum_i \theta_{iz}^{(U)}}, \tag{10}$$

by differentiating equation (8) with respect to $\theta_{iz}^{(U)}$ and $\theta_{jz}^{(V)}$. In statistics, this is known as ECM [29]. The criterion that vertex $i$ in $U$ belongs to in module $z$ is $\sum_j A_{ij} q_{ij}(z) \geqslant 1$ and that of vertex $j$ in $V$ in module z is $\sum_i A_{ij} q_{ij}(z) \geqslant 1$.

## 2.3. Mixture network as a unified framework

As noted, the mixture network can be considered as a general case, with the unipartite and bipartite networks as limiting cases. Consequently, we can define the generative model in a unified framework.

Given a graph $G = (U, V, E)$, where $U$ and $V$ represent two sets of vertices and $E$ is the set of edges between vertices in $U$ and $V$, the relationship between $U$ and $V$ determines the type of $G$, which can classify networks into three types. First is the unipartite network: when $U$ and $V$ are identical, all edges are within the same set of vertices, then $G$ is unipartite. Second is the bipartite network: when $U$ and $V$ are disjointed, $G$ is bipartite and each edge in $E$ links nodes from two disjointed sets. Third is the mixture network: when $U$ and $V$ are neither identical nor disjointed, i.e., they have some common vertices but not all vertices are the same, it is a mixture network.

Vertex labels, or identifiers, are required to determine whether a vertex is shared by two vertex sets. Let $l_i^{(U)}$ denote the label of vertex $i$ in $U$, and $l_j^{(V)}$ denote the label of vertex $j$ in $V$. In each set, any label is unique. When the label for vertex $i$ in $U$ is the same as that of vertex $j$ in $V$, $i$ and $j$ are the same vertex rather than two different vertices. For simplicity, let $O_U$ and $O_V$ denote two sets of nodes in $U$ and $V$ that share common labels (common vertices), while $S_U$ and $S_V$ denote two sets of vertices in $U$ and $V$ that do not share any common labels (specific vertices).

An illustration of the relationship between the mixture network, the unipartite network, and the bipartite network is given in figure 1.

The generalization of the bipartite network version model to that of the mixture network is based on a simple idea: if two vertices are actually the same, they should have only one underlying tendency to belong to each module, which is reached by constraining each pair of tendencies of these two nodes to each module to be the same. This idea is quite similar to that in the comment of Costa and Hensen [31], which suggests that a unipartite network can be transformed into a bipartite network by adding explicit constraints that specify that each pair of corresponding nodes must belong to the same community. Again, the tendencies of vertex $i$ in $U$ and vertex $j$ in $V$ belonging to module z are denoted as $\theta_{iz}^{(U)}$ and $\theta_{jz}^{(V)}$. Consequently, the likelihood function of a mixture network is

$$P\left(G\middle|\theta^{(U)}, \theta^{(V)}\right) = \prod_{ij} \frac{\left(\sum_z \theta_{iz}^{(U)} \theta_{jz}^{(V)}\right)^{A_{ij}}}{A_{ij}!} \exp\left(-\sum_z \theta_{iz}^{(U)} \theta_{jz}^{(V)}\right), \qquad (11)$$

subject to constraints that

$$\delta_{l_i^{(U)}, l_j^{(V)}} \left(\theta_{iz}^{(U)} - \theta_{jz}^{(V)}\right) = 0 \qquad \forall\ i, j, z, \qquad (12)$$

where $\delta$ is the Kronecker delta function. Fundamentally speaking, a random network generated by this generative model is a directed network, because its unipartite part does not necessarily need to be symmetric[4]. For an undirected network, we adopt the view that an undirected edge can be viewed as two directed edges in opposite directions [32]. For self-loops, a slight difference exists compared to those in BKN. Unlike other undirected edges, a self-loop only has one corresponding element in the adjacency matrix. Thus, when vertex $i$ in $U$ and vertex $j$ in $V$ are actually one node and they have a link (i.e., self-loop), the value of $A_{ij}$ is 1 rather than 2, regardless of whether the network is directed or undirected.

Such an optimization problem with equality constraints can be solved using the Lagrange multiplier. (The proof is available in appendix A.)

We introduce an arbitrary variable $q_{ij}(z)$ that satisfies $\sum_z q_{ij}(z) = 1$. The optimal value of $q_{ij}(z)$ is given by

$$q_{ij}(z) = \frac{\theta_{iz}^{(U)} \theta_{jz}^{(V)}}{\sum_z \theta_{iz}^{(U)} \theta_{jz}^{(V)}}. \qquad (13)$$

For $\theta_{iz}^{(U)}$ and $\theta_{jz}^{(V)}$, two conditions will be considered separately:

---

[4] We also propose a model specific to a mixture undirected network in the supplementary information. However, we could not see any significant difference between the specific model and the model mentioned in the main text from simulation results.

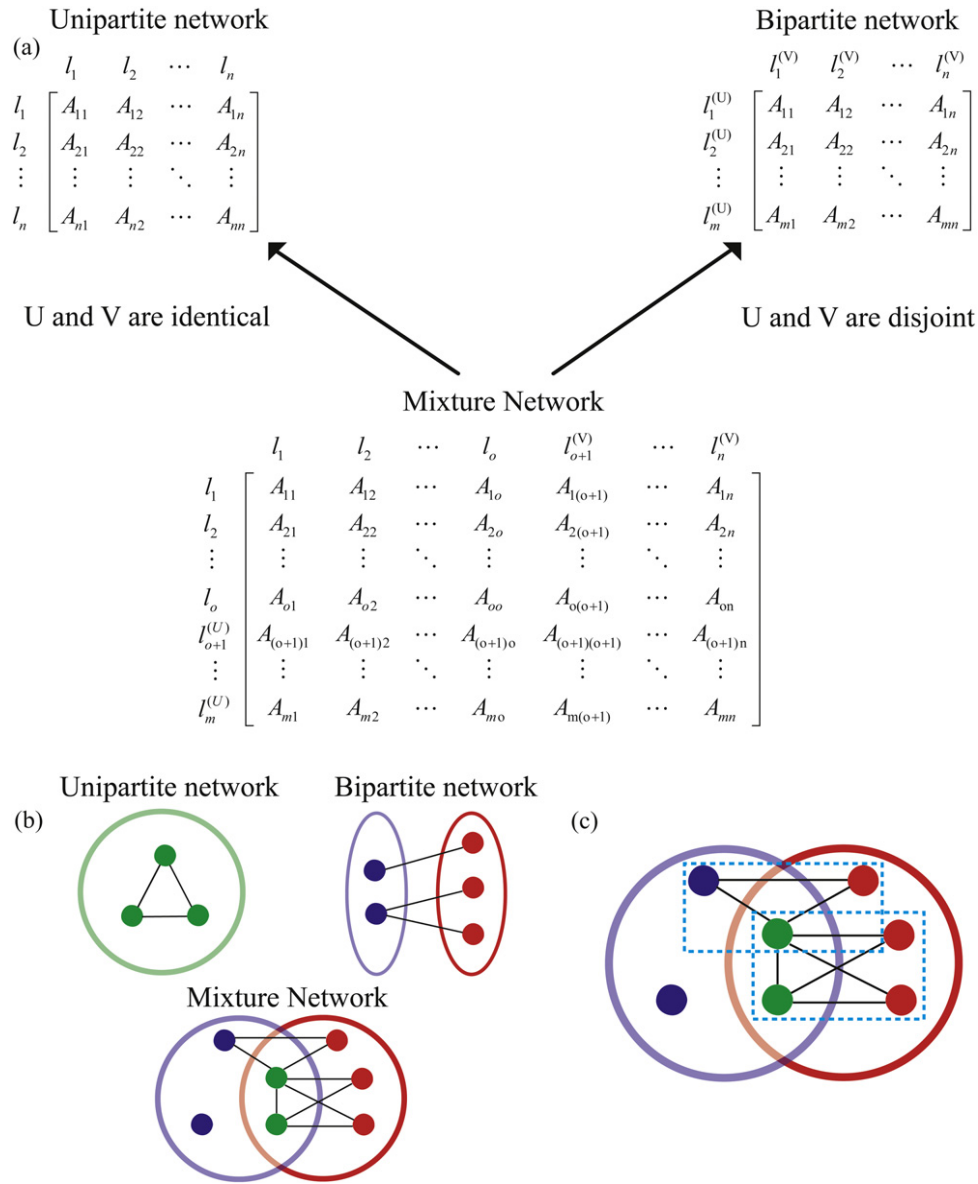

**Figure 1.** Schematic illustration of the concept of a mixture network. (a) The mixture network can be considered as a general case. The elements of the adjacency matrices are enclosed in square brackets and the labels of the vertices are placed at the left or top. For the convenience of the illustration, the adjacency matrix for the mixture network is rearranged so that the submatrix of all common vertices is in the upper left corner and the labels for each pair in this submatrix at the $i$th row and $i$th column both equal $l_i$. For a mixture network, when $U$ and $V$ are identical, only the submatrix of all common vertices remains, which is the adjacency matrix for unipartite. However, when $U$ and $V$ are disjointed, only the submatrix of all specific vertices remains, which is the adjacency matrix for bipartite. (b) Simple examples for three types of networks. The blue cycle denotes the vertex set for blue nodes, the red cycle denotes the set for red nodes, and the green nodes belong to both sets. (c) In community detection, a node can belong to one, many, or none of the modules (denoted by cyan dashed rectangles).

(1) For any pair of common vertices $i'$ in $O_U$ and $j'$ in $O_V$ that $\delta_{l_{i'}^{(U)}, l_{j'}^{(V)}} = 1$, we obtain

$$\theta_{i'z}^{(U)} = \frac{\sum_i A_{ij} q_{ij'}(z) + \sum_i A_{i'j} q_{ij}(z)}{\sum_i \theta_{iz}^{(U)} + \sum_j \theta_{jz}^{(V)}},$$

$$\theta_{j'z}^{(V)} = \frac{\sum_i A_{ij'} q_{ij'}(z) + \sum_i A_{ij} q_{ij}(z)}{\sum_i \theta_{iz}^{(U)} + \sum_j \theta_{jz}^{(V)}}; \tag{14}$$

(2) For the specific vertex $i$ in $S_U$ or $j$ in $S_V$, we obtain

$$\theta_{iz}^{(U)} = \frac{\sum_j A_{ij} q_{ij}(z)}{\sum_j \theta_{jz}^{(V)}}, \qquad \theta_{jz}^{(V)} = \frac{\sum_i A_{ij} q_{ij}(z)}{\sum_i \theta_{iz}^{(U)}}. \tag{15}$$

Our model for a mixture network is a general solution, because both BKN and the bipartite network version described in section 2.2 can be described as limiting cases. When all vertices of $U$ and $V$ are common vertices (unipartite network), only equation (14) remains. Because the network that BKN studied is symmetric, we have $\sum_i A_{ij} q_{ij}(z) = \sum_i A_{ij} q_{ij}(z)$ and $\sum_i \theta_{iz}^{(U)} = \sum_j \theta_{jz}^{(V)}$, and equation (14) can be reduced to equation (5). When all vertices of $U$ and $V$ are specific vertices (bipartite network), only equation (15) remains, which is the same as equation (10). Conversely, the form of equations for $q_{ij}(z)$ remains constant for all three versions of the models.

In the general model, the criteria for a node to belong to a module depend on whether this node is a common or specific vertex and whether the network is directed or undirected. For a specific vertex, the criterion is same as that of the bipartite network version model: vertex $i$ in $S_U$ belongs to module $z$ if $\sum_j A_{ij} q_{ij}(z) \geqslant 1$, and vertex j in $S_V$ belongs to module $z$ if $\sum_i A_{ij} q_{ij}(z) \geqslant 1$. For a common vertex, the criterion depends on whether the network is directed or undirected. When the network is undirected, the unipartite part of the network is symmetric. For a pair of corresponding common vertices $i'$ and $j'$, we only need to count the links between the common vertices once, i.e., $\sum_{i \in S_U} A_{ij'} q_{ij'}(z) + \sum_i A_{i'j} q_{ij}(z) \geqslant 1$. When the network is directed, links between the common vertices should be counted twice rather than once, i.e., $\sum_i A_{ij'} q_{ij'}(z) + \sum_j A_{i'j} q_{ij}(z) \geqslant 1$. According to the criteria, a node can belong to one module, several modules, or none of the modules.

## 3. Simulations of consistency tests

In this section, we perform several consistency tests that test the algorithm in terms of the random networks generated from the generative model itself with different parameters.

### 3.1. Synthetic networks generated by the generative model

Using the generative model described in section 2 to generate random networks can be viewed as the reverse process of finding network modules using the same model. When finding modules, we infer module structures from the adjacency matrix. When the model is used to generate random networks, community structures are known and adjacency matrices are generated based on the model's hidden variables.

The process of generating a random mixture network is described as follows: (1) assign module membership to vertices; (2) set expected number of links for nodes to each module; (3) calculate parameters; (4) calculate expected number of links between each pair of nodes; and (5) generate the adjacency matrix. In this study, we generate adjacency matrices with binary elements by forcing their elements to be 1 if they are larger than 1, despite that our model can deal with multi-edge problem technically.

Let $k_{iz}^{(U)} = \sum_j A_{ij} q_{ij}(z)$ and $k_{jz}^{(V)} = \sum_i A_{ij} q_{ij}(z)$, which are set by the user to denote the expected number of links belonging to module $z$ that link to vertex $i$ in $U$ and vertex $j$ in $V$, respectively. The values of $\sum_{i \in O_U} \theta_{iz}^{(U)}$, $\sum_{i \in S_U} \theta_{iz}^{(U)}$, $\sum_{j \in S_V} \theta_{jz}^{(V)}$, and $\sum_{j \in O_V} \theta_{jz}^{(V)}$ can be given by

$$\sum_{i \in O_U} \theta_{iz}^{(U)} = \sqrt{\frac{\left(\sum_{i \in O_U} k_{iz}^{(U)} + \sum_{j \in O_V} k_{jz}^{(V)}\right)^2 - \left(\sum_{i \in S_U} k_{iz}^{(U)} - \sum_{j \in S_V} k_{jz}^{(V)}\right)^2}{2\left(\sum_{i \in S_U} k_{iz}^{(U)} + \sum_{j \in S_V} k_{jz}^{(V)} + \sum_{i \in O_U} k_{iz}^{(U)} + \sum_{j \in O_V} k_{jz}^{(V)}\right)}},$$

$$\sum_{i \in S_U} \theta_{iz}^{(U)} = \frac{2 \sum_{i \in S_U} k_{iz}^{(U)}}{\sum_{j \in S_V} k_{jz}^{(V)} - \sum_{i \in S_U} k_{iz}^{(U)} + \sum_{i \in O_U} k_{iz}^{(U)} + \sum_{j \in O_V} k_{jz}^{(V)}} \sum_{i \in O_U} \theta_{iz}^{(U)},$$

$$\sum_{j \in S_V} \theta_{jz}^{(V)} = \frac{2 \sum_{j \in S_V} k_{jz}^{(V)}}{\sum_{i \in S_U} k_{iz}^{(U)} - \sum_{j \in S_V} k_{jz}^{(V)} + \sum_{i \in O_U} k_{iz}^{(U)} + \sum_{j \in O_V} k_{jz}^{(V)}} \sum_{i \in O_U} \theta_{iz}^{(U)},$$

$$\sum_{j \in O_V} \theta_{jz}^{(V)} = \sum_{i \in O_U} \theta_{iz}^{(U)}. \tag{16}$$

For common vertices $i'$ in $O_U$ and $j'$ in $O_V$ that $\delta_{l_i^{(U)}, l_j^{(V)}} = 1$,

$$\theta_{i'z}^{(U)} = \sum_{i \in O_U} \theta_{iz}^{(U)} \frac{k_{i'z}^{(U)} + k_{j'z}^{(V)}}{\sum_{i \in O_U} k_{iz}^{(U)} + \sum_{j \in O_V} k_{jz}^{(V)}},$$

$$\theta_{j'z}^{(V)} = \sum_{j \in O_V} \theta_{jz}^{(V)} \frac{k_{i'z}^{(U)} + k_{j'z}^{(V)}}{\sum_{i \in O_U} k_{iz}^{(U)} + \sum_{j \in O_V} k_{jz}^{(V)}}. \tag{17}$$

For specific vertices $i$ in $S_U$ and $j$ in $S_V$,

$$\theta_{iz}^{(U)} = \sum_{i \in S_U} \theta_{iz}^{(U)} \frac{k_{iz}^{(U)}}{\sum_{i \in S_U} k_{iz}^{(U)}}, \quad \theta_{jz}^{(V)} = \sum_{j \in S_V} \theta_{jz}^{(V)} \frac{k_{jz}^{(V)}}{\sum_{j \in S_V} k_{jz}^{(V)}}. \tag{18}$$

The process for generating a mixture network, as described here, can also be used to generate a unipartite directed network. However, for our bipartite network version, the model identifiability problem should be addressed before the model is used to generate random bipartite networks (the solution to parameters is not unique, which is treated in detail in appendix B and discussed in section 8). This problem can be easily solved by introducing an extra constraint $\sum_i \theta_{iz}^{(U)} = \sum_j \theta_{jz}^{(V)}$. Consequently, we have

$$\theta_{iz}^{(U)} = \frac{k_{iz}^{(U)}}{\sqrt{\sum_i k_{iz}^{(U)}}}, \ \ \theta_{jz}^{(V)} = \frac{k_{jz}^{(V)}}{\sqrt{\sum_j k_{jz}^{(V)}}}. \tag{19}$$

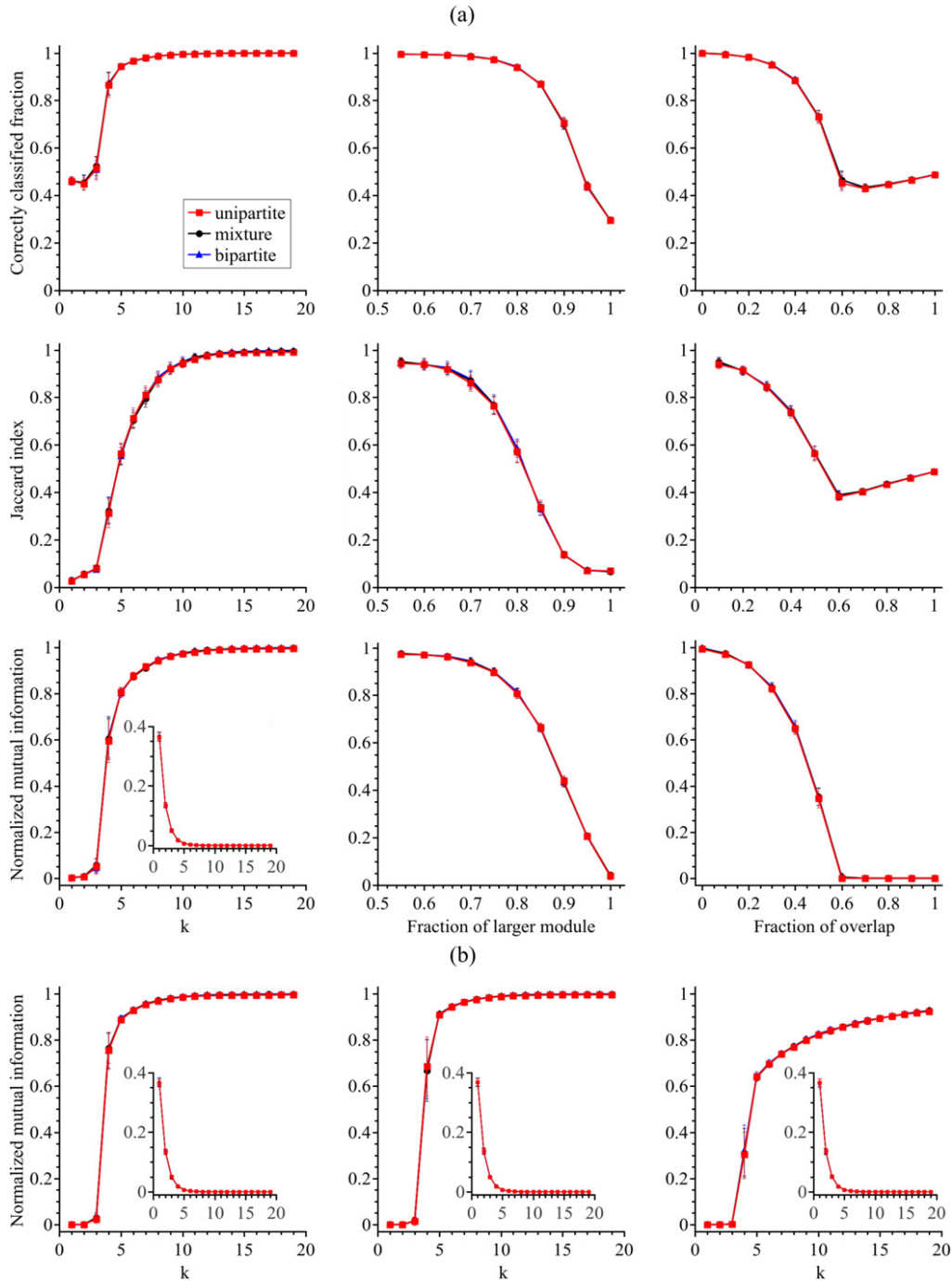

**Figure 2.** Results from consistency tests described in the text for random networks (a) with two modules and (b) with more than two modules. Each point is the average over 50 random networks. For each network, 30 random initializations are used, and the run with the highest log likelihood is chosen as the result. The insets show the fraction of isolated nodes for networks of each $k$ for the corresponding tests. For the second and the third sets of testing of random networks with two modules, because $k$ is fixed at a large value (10), only a few networks occasionally contain isolated node(s) (fractions are close to 0), the effects of which on the measured indices are negligible. Thus, their fractions of isolated nodes are not shown.

The details are available in appendix B.

### 3.2. Simulation results

We use three indices to measure the performance of the algorithm in this subsection: (1) the fraction of correctly classified vertices, (2) the Jaccard index for overlapping vertices (vertices belonging to more than one module), and (3) variant normalized mutual information (NMI) [33]. Because our algorithm allows one vertex to belong to multiple modules, a node is correctly classified if the predicted memberships for all modules are correct. The definition of the Jaccard index is as follows: let $S_1$ denote the known set of overlapping vertices and $S_2$ is the predicted set of overlapping vertices—the Jaccard index is defined as $|S_1 \cap S_2|/|S_1 \cup S_2|$. NMI measures the similarity between the actual community structures and the predictions from the perspective of information theory: it is one if the prediction is the same as the actual condition and zero when they are independent. If a generated random network contains any isolated nodes, they would be removed from the network before detecting the modules, and therefore cannot be used to measure the performance of the algorithm.

The total number of vertices is fixed to 2000 for all networks in our consistency tests. For random unipartite networks, adjacency matrices of $2000 \times 2000$ are generated. For random bipartite networks, adjacency matrices of $1000 \times 1000$ are generated so that the total number of vertices is unchanged. For mixture networks, adjacency matrices of $1500 \times 1500$ are generated with 1000 common vertices (50%); the total number of unique vertices is still 2000.

First, we perform three consistency tests on random networks with two modules. In the first sets of consistency testing, 55% of vertices belong to module 1 and 55% to module 2 (10% to both modules). The total expected number of links of each vertex is set to be $k$. That is, if a node is a specific vertex that only belongs to one module, its expected number of linkages to this module is $k$; on the other hand, if a specific vertex belongs to two modules, its expected number of linkages to both modules is $k/2$. For a common vertex, the number of linkages in both vertex sets is $k/2$ if it belongs to only one module and $k/4$ if it belongs to two modules.

We test the performance of the algorithm by varying $k$. When $k$ is small, the actual modular structures of the networks are quite hard to find. When $k$ is large, it is quite easy to reveal the modules of networks. As shown in the left panels of figure 2(a), the performance of the algorithm on all types of networks increases with $k$ as expected. When $k \leqslant 3$, the fractions of correctly classified vertices are far away from zero, but NMIs are quite close to zero, indicating that predictions for networks with low degrees are random with respect to the actual community structures. The performance at $k = 4$ increases rapidly. When $k \geqslant 10$, the algorithm can determine the modular structures for all three types of networks with high performance.

In the second set of consistency testing, $k$ is fixed at 10, the fraction of overlapping vertices remains at 10%, and the fraction of vertices belonging to module 1 increases. The condition when the fraction of vertices belonging to module 1 is smaller than the fraction belonging to module 2 does not need to be discussed because it is equivalent to the condition considered here. As suggested by the middle panels of figure 2(a), all indices decrease with further unbalanced fractions of two modules.

In the third set of testing, $k$ is fixed at 10, the fractions of modules 1 and 2 are set to be equal, and the fraction of overlapping vertices increases. The right panels of figure 2(a) indicate that all indices decrease with an increase of the fraction of overlapping vertices ($\leqslant$0.6). When the fraction of overlapping vertices is large, the fractions of correctly classified vertices and the Jaccard indices increase, but the NMIs are close to zero, indicating that the predictions for networks with large fractions of overlapping vertices are also random with respect to the actual community structures.

We also perform three consistency tests on random networks of more than two modules. In these tests, the total number of vertices is still fixed to 2000, the fraction of overlapping nodes is set to 10%, and the fractions of nodes specifically belonging to each module are equal but $k$ varies. For the overlapping nodes, the number of modules they belong to is restricted by an integer interval given by the user. Once the interval is given, we randomly generate the number of modules for each overlapping node and randomly assign it to a number of modules.

First, we set the number of modules to be five and each overlapping node belongs to two modules. Because the fraction of correctly classified vertices is difficult to calculate for random networks with more than two modules, and the Jaccard index for overlapping vertices cannot reflect whether they are assigned to the correct modules, these two indices are not calculated in the following tests. As shown in the left panel of figure 2(b), NMIs increase with increasing $k$. Similar results are observed for two sets of random networks with 10 modules, the first in which each overlapping vertex belongs to two modules (figure 2(b), middle panel) and the second of which belongs to 2–10 modules (figure 2(b), right panel). For the latter, the curves for all three types of networks increase more slowly than for the former.

In all sets of testing, the curves of the indices for all three types of networks overlap with each other very well at each point (figure 2). This suggests that our algorithm performs equally for all three types of networks generated by the same parameters.

In summary, the consistency test results suggest that our algorithm can perform well for a large range of parameters for all three types of networks. All indices increase with increasing $k$. For random networks with two modules, all indices approach one when the difference between the fractions of two modules or the fraction of overlapping vertices is small.

## 4. Simulations of synthetic networks generated by sampling

### 4.1. Synthetic networks generated by sampling

Next, we test the performance of the algorithm based on synthetic networks generated by sampling, the pseudocode of which is given in figure 3. In this process, $r$ random unipartite networks are generated for each measured parameter of network quality. And for each random unipartite network, we generate $s$ networks by sampling. Note that in sampling, some information is lost, because many elements in the adjacency matrix of the unipartite network become unknown in sampled networks.

Random unipartite networks used in this section are generated using two well-known benchmarks that are of nonoverlapping community: the Girvan–Newman (GN) benchmark [2, 34] and the Lancichinetti–Fortunato–Radicchi (LFR) benchmark [35–37]. Using algorithms that allow an overlapping community to detect a nonoverlapping community is straightforward by assigning each node to its most probable module. For the GN benchmark, there are 128 vertices and 4 modules in each generated random network, with an expected degree of 16 for

```
foreach network quality
  (1) generate r random unipartite networks;
  foreach random unipartite network
    foreach sampling rule
      (2) sample s sub-networks;
      (3) find modules;
      (4) calculate prediction measures;
    end
  end
end
```

**Figure 3.** Pseudocode of generating random networks by sampling.

each node. Compared to the GN benchmark, it is more difficult to examine community structure in random networks generated by the LFR benchmark. For the LFR benchmark, we generate random unipartite networks with 1000 nodes, with community size ranging from 20 to 100 (labeled 'B'), and all other parameters are the same as those used in [37]. Both directed and undirected random networks of the two benchmarks are generated using the software described in [35–37]. The difficulty of finding network modules is controlled by the mixing parameter $\mu$, which determines the ratio of the external degree linking to other modules to the total degree of a node. When $\mu$ increases, the community structure is more mixed. The community structure vanishes when $\mu$ is large enough.

We define two sampling rules: symmetric and asymmetric sampling. For symmetric sampling, the fractions of specific vertices in $U$ and $V$ are the same (denoted as $u$ and $v$), with the fraction of common vertices being changed (denoted as $c$). For both benchmarks, we change $c$ from 0 to 1, which determines the type of synthetic network generated by sampling: (1) $c = 0$, meaning the sampled networks are bipartite; (2) $c = 1$, meaning the sampled networks are unipartite and identical to the original unipartite network; and (3) $0 < c < 1$, meaning the sampled networks are mixture networks. For asymmetric sampling, we fix $c$ and then change $u$ and $v$ accordingly. These three parameters are not independent; their sum is always one.

### 4.2. Simulation results

First, we apply our algorithm to a set of symmetric sampling undirected networks generated from the GN benchmark. As shown in figure 4 (left upper panel), when $\mu$ increases, NMIs decrease for random networks with any $c$, as expected. When $\mu \leqslant 0.3$, there is no large difference between the NMIs of networks with different $c$. When $\mu = 0.4$, the performance for our model on networks with small $c$ begins to decrease. The smaller $c$ is, the worse the performance. It is not hard to understand, because the fraction of unknown edges increases with the decrease of $c$. When $\mu \geqslant 0.5$, the performance vanishes. The tendency for the performance of symmetric sampling for directed networks is generally similar to those of undirected ones, despite the difference in the point of decrease (figure 4, left lower panel).

For the GN benchmark, we also apply the algorithm to four sets of asymmetric sampling networks, $c$ of which are fixed to 0.2 and 0.8 (two tests for undirected networks and two for directed networks). For the former tests, the performance for networks of $u = 0.5$ decreases the slowest, because they have more remaining known edges than those of other fractions (figure 4, middle panels). For the latter tests, performances for all $u$ are quite similar because $c = 0.8$, and no matter how $u$ and $v$ change, the information loss is quite similar (figure 4, right panels).

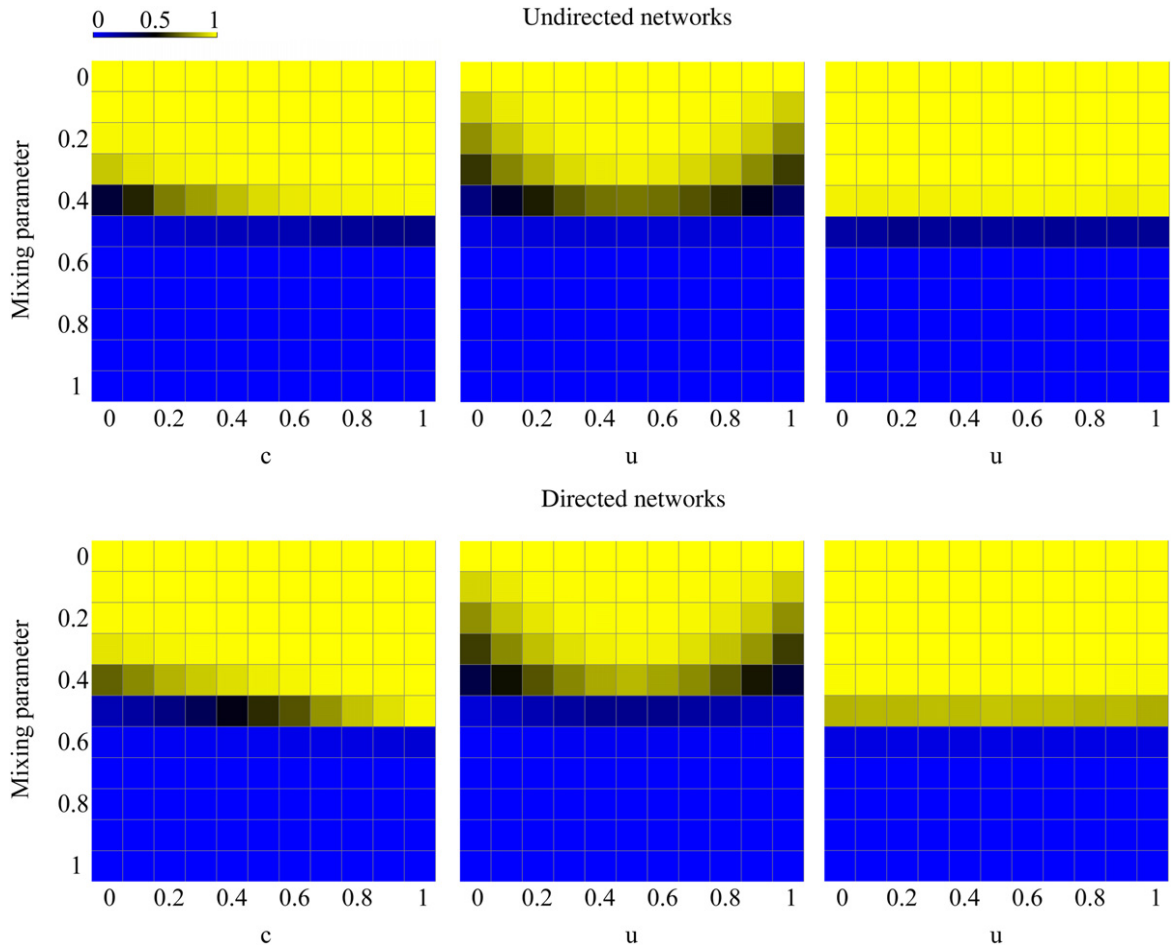

**Figure 4.** Heat maps for NMI of our model on random networks generated by symmetric sampling (left panels), asymmetric sampling with $c = 0.2$ (middle panels), and asymmetric sampling with $c = 0.8$ (right panels) using the GN benchmark (directed and undirected networks), which are created using matrix2png [38]. Both $r$ and $s$ are set to be 10. Fifty random initializations are used for the module detection of each network.

For the LFR benchmark, we also apply our algorithm to two sets of symmetric sampling networks and four sets of asymmetric sampling networks. Similar results are obtained from tests of this more difficult benchmark. For the symmetric sampling undirected networks (figure 5, left upper panel), when $\mu <= 0.5$, the algorithm easily finds the modules, despite that the performance for networks with $c$ close to 0 begins to decrease when $\mu$ is close to 0.5. When $\mu = 0.6$, the performance decreases. The smaller $c$ is, the worse the performance. When $\mu \geqslant 0.7$, the algorithm fails. For the remaining five tests, the performance tendency is generally similar to that of the GN benchmark, despite the difference in the point of decrease (figure 5, left bottom, middle, and right panels). The similar tendencies between the performance of our algorithm on the easier and more difficult tests suggest that our model is applicable not only to small and easy networks, but also to large and difficult ones.

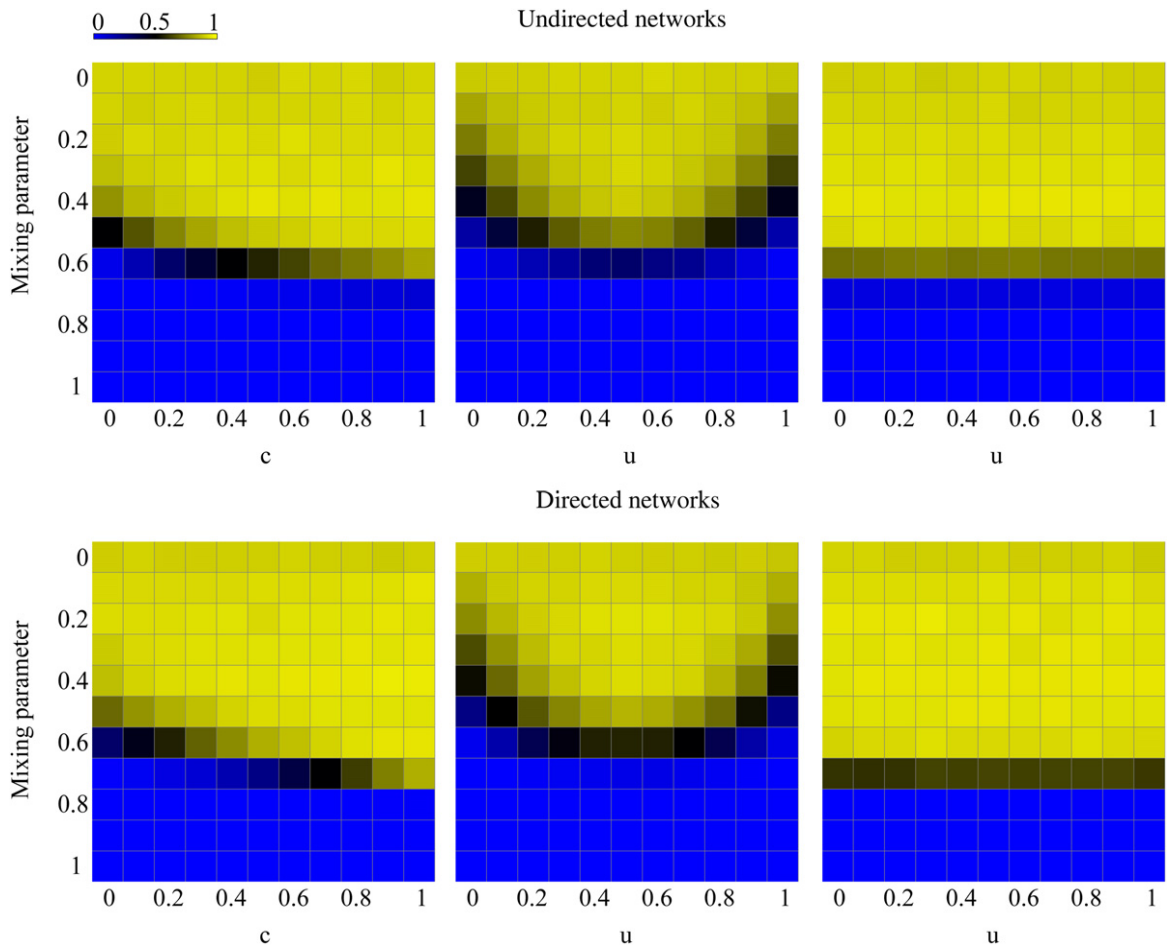

**Figure 5.** Heat maps for NMI of our model on random networks generated by symmetric sampling (left panels), asymmetric sampling with $c = 0.2$ (middle panels), and asymmetric sampling with $c = 0.8$ (right panels) using the LFR benchmark (directed and undirected networks), which are created using matrix2png. Here, $r$ is set to be 10 and $s$ is set to be 5. Ten random initializations are used for the module detection of each network.

## 5. Simulations of unipartite directed networks

In this section, we apply our model to random unipartite directed networks generated by GN and LFR benchmarks, and compare it with the optimization of a modularity defined in the context of a directed network (denoted as DM) [39]. The optimization process is completed using the RADATOOLS software [40] using extremal optimization [41].

The data shown in figure 6 indicate that our model has similar performance to DM on random unipartite directed networks generated using GN and LFR benchmarks. The number of modules found by DM tends to depart from the actual number when $\mu$ is large for both benchmarks (figure 6, insets).

We further optimize the results of DM and our model using a combinatorial optimization strategy of the fast algorithm [42] and the reposition algorithm using the RADATOOLS software. For the GN benchmark, the performances for the optimized algorithms are similar to those of nonoptimized algorithms (figure 6(a)). However, the performances for algorithms increase after optimization for the LFR benchmark (figure 6(b)).

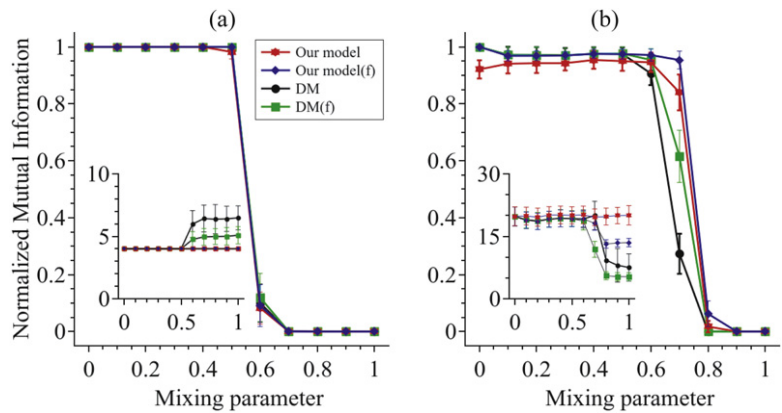

**Figure 6.** Performance of the algorithms on random unipartite directed networks generated using (a) GN and (b) LFR benchmarks. The optimized algorithms are marked with '(f)' following their names. Each data point is averaged over 100 networks. The insets show the number of modules given to our model (i.e., the actual number of modules) and those found by DM, DM(f), and our model(f). The number of random initializations of our model for (a) and (b) is 100 and 10, respectively.

## 6. Simulations of bipartite networks

Next, we apply our algorithm to a set of randomly generated bipartite networks, following [9]. Each random bipartite network contains 100 vertices and consists of 5 modules. In each module, there are 12 vertices in $U$ and 8 vertices in $V$. The difficulty of prediction is controlled by the ratio $p_{out}/p_{in}$, in which $p_{out}$ denotes the probability of vertices being connected between different modules and $p_{in}$ denotes the probability of vertices being linked within the same module. Links only exist between vertices in $U$ and $V$.

We compare our algorithm with the label propagation algorithm for bipartite networks (LPAb) [11] and a spectral algorithm based on singular value decomposition [12]; both of which are state-of-the-art algorithms that can find community structure in bipartite networks. LPAb is based on a label propagation algorithm (LPA) [43], which assigns unique labels to nodes and repeatedly updates the label of each vertex by assigning the most frequent labels of its neighbors until it meets the terminal condition. Barber and Clark [11] reformulated LPA as an optimization problem, addressed its drawbacks with additional constraints, and produced several variant LPA algorithms. LPAb is one of the variants that can be used to find modules in bipartite networks. LPA and its variants can determine the number of modules by themselves.

For bipartite networks, the above mentioned spectral algorithm finds the modules in three stages: (1) matrix factorization of the adjacency matrix using singular value decomposition; (2) dimensionality reduction using $k$-rank least-squares approximation; and (3) clustering vertices in the reduced space. In the third step, various clustering methods can be used, including the $k$-means clustering algorithm and the cosine matrix reordering mentioned in [12]. Here, we used $k$-means for the spectral algorithm, denoted as SVD ($k$-means).

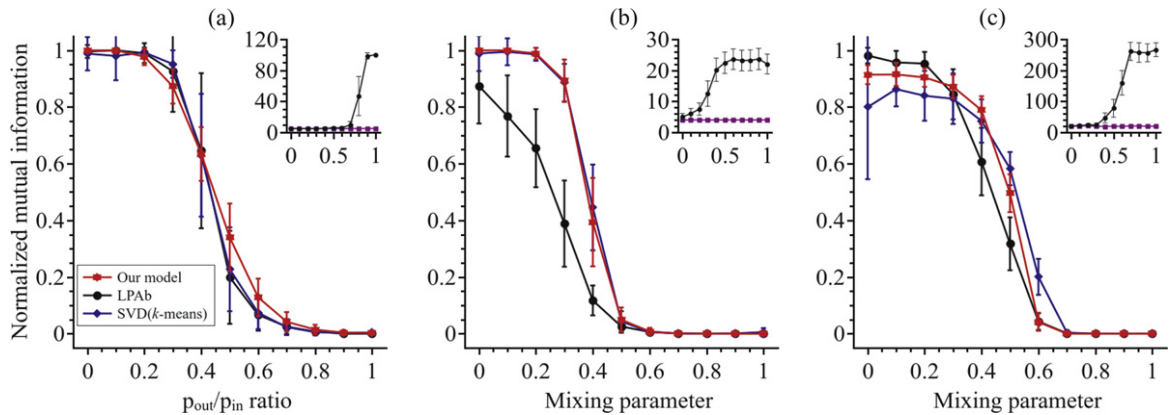

**Figure 7.** Performance of the algorithms on random bipartite networks (a) generated following [9] averaged over 100 networks for each data point, (b) generated by symmetric sampling from the GN benchmark ($r = 10$, $s = 10$), and (c) generated by symmetric sampling from the LFR benchmark ($r = 10$, $s = 5$). The insets show the actual number of modules (violet solid line, rhombus) and those found by LPAb (black solid line, ellipse). For SVD ($k$-means) and our model, the number of modules is given. The number of random initializations of our model for (a–c) is 100, 50, and 10, respectively.

The data shown in figure 7(a) suggest that all three algorithms have similar performance in this set of tests (100-node randomly generated bipartite networks). Notably, when $p_{out}/p_{in}$ is small, LPAb gives the correct number of modules and when $p_{out}/p_{in}$ is large, it gives a very large number of modules (figure 7(a), inset).

We also compare our algorithm with LPAb and SVD ($k$-means) for bipartite networks generated by symmetric sampling. For bipartite networks sampled from the GN benchmark, our algorithm and SVD ($k$-means) perform better than LPAb when $\mu \leqslant 0.4$ (figure 7(b)). A possible explanation for why the performance of LPAb is not as good in this dataset is that the number of modules given by the LPAb departs from the correct number, even when $\mu$ is small [figure 7(b), inset]. For bipartite networks sampled from the LFR benchmark, all three algorithms have similar performance (figure 7(c)). An exception is for SVD ($k$-means) at $\mu = 0.0$. At this point, SVD ($k$-means) performs poorly and has a larger standard deviation. This phenomenon is caused by $k$-means clustering rather than singular value decomposition, because replacing $k$-means with hierarchical clustering [44] as the clustering method for the spectral algorithm can achieve good performance ($NMI = 1\pm0$) at $\mu = 0.0$.

Overall, the comparison of the three models indicates that the performance of our model, when used to detect nonoverlapping communities for bipartite networks, is competitive with LPAb and SVD ($k$-means).

## 7. Real networks

### 7.1. Southern women bipartite network

We apply our algorithm to a well-known real-world bipartite network: the southern women dataset [45], which consists of 18 women (W1–W18) and 14 social events (E1–E14). This dataset has been studied extensively to investigate the results of various methods [46].

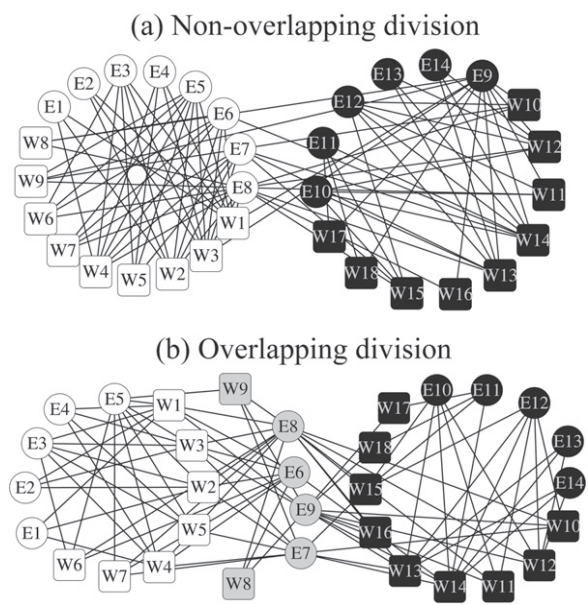

**Figure 8.** (a) Nonoverlapping and (b) overlapping division of the bipartite southern women network. Network visualization is created using Cytoscape [47]. Nodes colored black or white belong to only one module and those colored gray belong to both modules simultaneously.

To compare the result to those of other algorithms, we first display the result of our algorithm by assigning nodes to two nonoverlapping modules. The nonoverlapping partition divides the 18 women into W1–W9 and W10–W18, which agrees with the 'perfect' partition as suggested in [46] (figure 8(a)). Unlike many algorithms mentioned in [46], our algorithm assigns social events to modules along with women. For the social events, E1–E8 belong to the same module as W1–W9 and E9–E14 as W10–W18.

When allowing overlapping modules, W8, W9, and E6–E9 belong to both modules (figure 8(b)). This is quite reasonable, because all links only exist between those overlapping nodes and the remaining specific nodes of two modules. These overlapping nodes maintain the connection between these two modules.

## 7.2. Transcriptional regulatory mixture network

Next, we apply our algorithm to a recently released human transcriptional regulatory network[5] obtained from [15], which is a mixture directed network. This dataset consists of 82 TFs and 3998 downstream target genes, which include 31 TFs. The number of modules is set at 30. One hundred random initializations are used for the algorithm and the run with the highest log likelihood is chosen as the result. For 30 modules, the average number of TFs is 5.3 and the average number of target genes is 242.2.

Modular structures identified in the network can find genes that are known to be related to one another or offer new biological insights. The underlying assumption is that genes in the same module tend to be involved in the same or similar biological pathways. Gene ontology (GO) [48] is a standard tool used to associate biological functions with gene sets. Here, we perform GO enrichment analysis [49] to determine the enriched GO terms for downstream target genes. We report two case studies on the modules found in this transcriptional regulatory network. Supplementary figure 1 displays a module consisting of 3 TFs and 469 downstream target genes (including 2 TFs). The enriched GO terms are mainly related to the cell cycle and the metabolic process (supplementary table 1). There are three TFs here: *HDAC*2, *NFKB*1, and *NFYA*. All three genes are known to be related to the cell cycle [50–53] and *NFYA* is also related to the metabolic process [54].

---

[5] This dataset is known as 'enets7.K562_proximal_filtered_network.txt', which is described in detail in the supplementary information of Gerstein *et al*.

A second case study focuses on a module related to the immune system (supplementary figure 2). It consists of 130 downstream genes (including 2 TFs) and 10 TFs, 6 of which (*IRF*1, *POLR*3A, *POU*2F2, *PRDM*1, *FOSL*2, and *IRF*4) are related to the immune system [54–59]. For downstream genes, the enriched GO terms are shown in supplementary table 2, many of which are immune related. We predict that another 4 TFs (*PBX*3, *POU*5F1, *PPARGC*1A, and *MAFK*) may also be related to the immune system.

## 8. Discussion

In this paper, we present a probabilistic method for identifying community structures in unipartite, bipartite, and mixture networks in a unified framework. Our model can assign vertices to one module, multiple modules, or none of the modules. When the network is bipartite or a mixture, it can simultaneously assign memberships to both sets of nodes.

We test the performance of our algorithm by applying it to several sets of synthetic networks (both overlapping and nonoverlapping communities) and two real-world networks. Our model can perform well on large parameter ranges in synthetic networks and is competitive with other algorithms for unipartite or bipartite networks. In addition, our model is applicable to real-world networks. For the southern women bipartite dataset, our model offers reasonable overlapping and nonoverlapping community divisions. For the human transcriptional regulatory network (a mixture directed network), modules associated with biological functions are reported in two case studies, which can reveal related genes or provide new clues. Along with the simulation and real network results of BKN on unipartite undirected networks [6], and the additional algorithm performance comparison of BKN[6] with other algorithms that can identify modules in networks with overlapping communities for unipartite undirected networks [60], we conclude that our algorithm performs well for all three types of network and is competitive with other algorithms for unipartite or bipartite networks.

The main contributions of our model are as follows. First, it introduces the concept of the mixture network, which can represent a wide range of networks that are neither unipartite nor bipartite in many fields such as biology and social science, and can operate on this type of data. To our knowledge, this is the first network community detection algorithm to address this question. Second, it offers a unified framework for unipartite, bipartite, and mixture networks.

---

[6] BKN is known as the Poisson community model (POI) in Gopalan and Blei's paper. Gopalan and Blei compared their stochastic inference algorithm (SVI) with POI, INFOMAP (INF), and COPRA (COP) for finding overlapping communities in synthetic datasets of unipartite undirected networks. SVI perform as well as POI, both of which outperform INF and COP. Our model is degenerate to BKN (POI) when it is used to find modules for unipartite undirected networks.

Finally, this framework can potentially be used for the generalization of other community detection algorithms and other models in the fields of complex networks.

One may note that, for the bipartite network version of the model, the solution for $\theta_{iz}^{(U)}$ and $\theta_{jz}^{(V)}$ is not unique even for the same local maxima, that is, it suffers from the problem of model identifiability. For any solution, multiplying a positive constant $C$ for all $\theta^{(U)}$, and dividing the same constant for all $\theta^{(V)}$, is still a solution to the model. Nevertheless, because the product of $\theta_{iz}^{(U)}$ and $\theta_{jz}^{(V)}$ remains unchanged, which does not affect the value of $q_{ij}(z)$ (the quantity that we are concerned with), we can ignore this problem when the algorithm is used to detect network modules. However, when the method is used to generate random bipartite networks (section 3), a constraint for parameters is needed, as shown in appendix B. The model identifiability problem does not exist when the network is unipartite or a mixture.

Similar to [6], the main drawback of this algorithm is that it is unable to determine the number of module $K$ from the data. Methods for model selection, such as the Akaike information criterion [61], the Bayesian information criterion [62], or the likelihood ratio test [63], are not applicable here for the same reason as in [6], i.e., many parameters are zero which violates the assumption of model selection methods.

We also note that in real-world data, edges are sometimes associated with weights or scores other than Poisson that can be used to indicate the existence or reliability of the links [15, 19, 21]. We plan to address this issue, combined with the problem of determining the number of modules, in future studies.

## Acknowledgments

The authors gratefully acknowledge Junwei Wang for useful and detailed discussions on the statistical techniques related to the model and careful reading of the manuscript; Ills Farkas, Minghua Deng, and Lin Wang for comments and suggestions; Iain Bruce and Xiang Liu for detailed proofreading of the manuscript; Jun Wang for checking the mathematics; and Lingli Jiang for her epistatic mini-array profile experiments that motivated this model. This work was supported in part by the National Natural Science Foundation of China.

## Appendix A. Solution of mixture network version model

The likelihood function and constraint are given by equations (11) and (12). Taking the logarithm of equation (11), and introducing an arbitrary variable $q_{ij}(z)$ that satisfies $\sum_z q_{ij}(z) = 1$, we obtain

$$\ln P\left(G\big|\theta^{(U)}, \theta^{(V)}\right) \geqslant \sum_{ijz}\left[A_{ij}q_{ij}(z)\ln\left(\frac{\theta_{iz}^{(U)}\theta_{jz}^{(V)}}{q_{ij}(z)}\right) - \theta_{iz}^{(U)}\theta_{jz}^{(V)}\right]. \qquad (A.1)$$

Now, we consider the constraint in equation (12). Let $L$ denote the target function, which is given by

$$L = \sum_{ijz} \left[ A_{ij} q_{ij}(z) \ln\left( \frac{\theta_{iz}^{(U)} \theta_{jz}^{(V)}}{q_{ij}(z)} \right) - \theta_{iz}^{(U)} \theta_{jz}^{(V)} \right] + \sum_{ijz} c_{ijz} \delta_{l_i^{(U)}, l_j^{(V)}} \left( \theta_{iz}^{(U)} - \theta_{jz}^{(V)} \right), \qquad \text{(A.2)}$$

where $c_{ijz}$ is the Lagrange multiplier. Differentiating equation (A.2), with respect to $\theta_{iz}^{(U)}$, leads to

$$\begin{aligned} \frac{\partial L}{\partial \theta_{iz}^{(U)}} &= \sum_j \left[ \frac{A_{ij} q_{ij}(z)}{\theta_{iz}^{(U)}} - \theta_{jz}^{(V)} + c_{ijz} \delta_{l_i^{(U)}, l_j^{(V)}} \right] \\ &= \frac{\sum_j A_{ij} q_{ij}(z)}{\theta_{iz}^{(U)}} - \sum_j \theta_{jz}^{(V)} + \sum_j c_{ijz} \delta_{l_i^{(U)}, l_j^{(V)}} \\ &= 0. \end{aligned} \qquad \text{(A.3)}$$

Consequently,

$$\theta_{iz}^{(U)} = \frac{\sum_j A_{ij} q_{ij}(z)}{\sum_j \theta_{jz}^{(V)} - \sum_j c_{ijz} \delta_{l_i^{(U)}, l_j^{(V)}}}. \qquad \text{(A.4)}$$

Similarly, we have

$$\theta_{jz}^{(U)} = \frac{\sum_i A_{ij} q_{ij}(z)}{\sum_i \theta_{iz}^{(U)} + \sum_i c_{ijz} \delta_{l_i^{(U)}, l_j^{(V)}}}. \qquad \text{(A.5)}$$

We shall only consider the condition for common vertices, because for other vertices, equations (A.4) and (A.5) become equation (10). For each pair of common vertices $i'$ and $j'$, $\delta_{l_{i'}^{(U)}, l_{j'}^{(V)}} = 1$. Inserting equations (A.4) and (A.5) into equation (12), we obtain

$$c_{i'j'z} = \frac{-\sum_i \theta_{iz}^{(U)} \sum_j A_{i'j} q_{i'j}(z) + \sum_j \theta_{jz}^{(V)} \sum_i A_{ij'} q_{ij'}(z)}{\sum_i A_{ij'} q_{ij'}(z) + \sum_j A_{i'j} q_{i'j}(z)} \qquad \forall\, z. \qquad \text{(A.6)}$$

Inserting equation (A.6) into equation (A.4),

$$\begin{aligned} \theta_{i'z}^{(U)} &= \frac{\sum_j A_{i'j} q_{i'j}(z) \sum_i A_{ij'} q_{ij'}(z) + \left( \sum_j A_{i'j} q_{i'j}(z) \right)^2}{\sum_j \theta_{jz}^{(V)} \sum_i A_{ij'} q_{ij'}(z) + \sum_i \theta_{iz}^{(U)} \sum_j A_{i'j} q_{i'j}(z)} \\ &= \frac{\sum_i A_{ij'} q_{ij'}(z) + \sum_j A_{i'j} q_{i'j}(z)}{\sum_j \theta_{jz}^{(V)} + \sum_i \theta_{iz}^{(U)}}. \end{aligned} \qquad \text{(A.7)}$$

Similarly, by inserting equation (A.6) into equation (A.5), we obtain

$$\theta_{j'z}^{(V)} = \frac{\sum_i A_{ij'} q_{ij'}(z) + \sum_j A_{i'j} q_{i'j}(z)}{\sum_i \theta_{iz}^{(U)} + \sum_j \theta_{jz}^{(V)}}. \qquad \text{(A.8)}$$

## Appendix B. Generating random networks

We start from equation (14). By multiplying the denominator of the right side at both sides of the first equation, and summing over all common vertices, we obtain

$$\sum_{j \in O_V} \theta_{jz}^{(V)} \sum_i \theta_{iz}^{(U)} + \sum_{i \in O_U} \theta_{iz}^{(U)} \sum_j \theta_{jz}^{(V)} = \sum_{i \in O_U} k_{iz}^{(U)} + \sum_{j \in O_V} k_{jz}^{(V)}. \tag{B.1}$$

Similarly, starting from equation (15), we get

$$\sum_{i \in S_U} \theta_{iz}^{(U)} \sum_j \theta_{jz}^{(V)} = \sum_{i \in S_U} k_{iz}^{(U)},$$
$$\sum_{j \in S_V} \theta_{jz}^{(V)} \sum_i \theta_{iz}^{(U)} = \sum_{j \in S_V} k_{jz}^{(V)}. \tag{B.2}$$

The goal is to determine the solutions for $\theta^{(U)}$ and $\theta^{(V)}$, given that each $k_{iz}^{(U)}$ and $k_{jz}^{(V)}$ is known. To make the symbols simpler, we define $X = \sum_{i \in S_U} \theta_{iz}^{(U)}$, $Y = \sum_{i \in O_U} \theta_{iz}^{(U)}$, $Z = \sum_{j \in S_V} \theta_{jz}^{(V)}$, and $W = \sum_{j \in O_V} \theta_{jz}^{(V)}$ so that

$$X + Y = \sum_i \theta_{iz}^{(U)}, \qquad Z + W = \sum_j \theta_{jz}^{(V)}, \tag{B.3}$$

and $\alpha = \sum_{i \in S_U} k_{iz}^{(U)}$, $\beta = \sum_{i \in O_U} k_{iz}^{(U)}$, $\gamma = \sum_{j \in S_V} k_{jz}^{(V)}$, and $\zeta = \sum_{j \in O_V} k_{jz}^{(V)}$, so that

$$\alpha + \beta = \sum_i k_{iz}^{(U)}, \qquad \gamma + \zeta = \sum_j k_{jz}^{(V)}. \tag{B.4}$$

Equations (B.1) and (B.2) can be rewritten as

$$W(X + Y) + Y(Z + W) = \beta + \zeta, \tag{B.5}$$

$$X(Z + W) = \alpha, \tag{B.6}$$

$$Z(X + Y) = \gamma. \tag{B.7}$$

Solving the equations using $Y = W$, we obtain

$$X = \frac{2\alpha}{\gamma - \alpha + \beta + \zeta} Y, \tag{B.8}$$

$$Z = \frac{2\gamma}{\alpha - \gamma + \beta + \zeta} Y. \tag{B.9}$$

Inserting equations (B.8) and (B.9) into equation (B.6), we get

$$Y = \sqrt{\frac{(\beta + \zeta)^2 - (\alpha - \gamma)^2}{2(\alpha + \gamma + \beta + \zeta)}}. \tag{B.10}$$

Now, the values of $\sum_{i \in O_U} \theta_{iz}^{(U)}$, $\sum_{i \in S_U} \theta_{iz}^{(U)}$, $\sum_{j \in S_V} \theta_{jz}^{(V)}$, and $\sum_{j \in O_V} \theta_{jz}^{(V)}$ can be given by

$$\sum_{i \in O_U} \theta_{iz}^{(U)} = \sqrt{\frac{\left(\sum_{i \in O_U} k_{iz}^{(U)} + \sum_{j \in O_V} k_{jz}^{(V)}\right)^2 - \left(\sum_{i \in S_U} k_{iz}^{(U)} - \sum_{j \in S_V} k_{jz}^{(V)}\right)^2}{2\left(\sum_{i \in S_U} k_{iz}^{(U)} + \sum_{j \in S_V} k_{jz}^{(V)} + \sum_{i \in O_U} k_{iz}^{(U)} + \sum_{j \in O_V} k_{jz}^{(V)}\right)}},$$

$$\sum_{i \in S_U} \theta_{iz}^{(U)} = \frac{2\sum_{i \in S_U} k_{iz}^{(U)}}{\sum_{j \in S_V} k_{jz}^{(V)} - \sum_{i \in S_U} k_{iz}^{(U)} + \sum_{i \in O_U} k_{iz}^{(U)} + \sum_{j \in O_V} k_{jz}^{(V)}} \sum_{i \in O_U} \theta_{iz}^{(U)},$$

$$\sum_{j \in S_V} \theta_{jz}^{(V)} = \frac{2\sum_{j \in S_V} k_{jz}^{(V)}}{\sum_{i \in S_U} k_{iz}^{(U)} - \sum_{j \in S_V} k_{jz}^{(V)} + \sum_{i \in O_U} k_{iz}^{(U)} + \sum_{j \in O_V} k_{jz}^{(V)}} \sum_{i \in O_U} \theta_{iz}^{(U)},$$

$$\sum_{j \in O_V} \theta_{jz}^{(V)} = \sum_{i \in O_U} \theta_{iz}^{(U)}. \tag{B.11}$$

For common vertices, starting from equation (14), we have

$$\sum_i \theta_{iz}^{(U)} + \sum_j \theta_{jz}^{(V)} = \frac{\sum_{i \in O_U} k_{iz}^{(U)} + \sum_{j \in O_V} k_{jz}^{(V)}}{\sum_{i \in O_U} \theta_{iz}^{(U)}}. \tag{B.12}$$

Inserting back into equation (14), we obtain

$$\theta_{i'z}^{(U)} = \sum_{i \in O_U} \theta_{iz}^{(U)} \frac{k_{i'z}^{(U)} + k_{j'z}^{(V)}}{\sum_{i \in O_U} k_{iz}^{(U)} + \sum_{j \in O_V} k_{jz}^{(V)}},$$

$$\theta_{j'z}^{(V)} = \sum_{j \in O_V} \theta_{jz}^{(V)} \frac{k_{i'z}^{(U)} + k_{j'z}^{(V)}}{\sum_{i \in O_U} k_{iz}^{(U)} + \sum_{j \in O_V} k_{jz}^{(V)}}. \tag{B.13}$$

Similarly, for specific nodes, we have

$$\theta_{iz}^{(U)} = \sum_{i \in S_U} \theta_{iz}^{(U)} \frac{k_{iz}^{(U)}}{\sum_{i \in S_U} k_{iz}^{(U)}}, \qquad \theta_{jz}^{(V)} = \sum_{j \in S_V} \theta_{jz}^{(V)} \frac{k_{jz}^{(V)}}{\sum_{j \in S_V} k_{jz}^{(V)}}. \tag{B.14}$$

Next, we will show that this solution agrees with that of BKN in the context of a unipartite undirected network. For such a network, $\alpha = 0$, $\gamma = 0$, and $\beta = \zeta$, as a consequence, $X = 0$, $Z = 0$, and

$$
\begin{aligned}
Y &= \sqrt{\frac{(\beta + \zeta)^2}{2(\beta + \zeta)}} \\
&= \sqrt{\frac{\beta + \zeta}{2}} \\
&= \sqrt{\beta} \\
&= \sqrt{\sum_{i \in O_U} k_{iz}^{(U)}},
\end{aligned}
\tag{B.15}
$$

which is exactly the same as equation (7) in [6].

On the other side, for bipartite networks, we have $\beta = 0$ and $\zeta = 0$ and therefore, $\alpha = \gamma$. Because the bipartite network version model suffers from the model identifiability problem, we can introduce an extra constraint $X = Z$ to solve the problem and obtain $X = Z = \sqrt{\alpha}$. Consequently,

$$
\theta_{iz}^{(U)} = \frac{k_{iz}^{(U)}}{\sqrt{\sum_i k_{iz}^{(U)}}}, \qquad \theta_{iz}^{(U)} = \frac{k_{iz}^{(U)}}{\sqrt{\sum_i k_{iz}^{(U)}}}.
\tag{B.16}
$$